\documentclass{JHEP3}
\usepackage{feynmp}

\title{Some Comments on an MeV Cold Dark Matter Scenario}

\author{Chen Jacoby and Shmuel Nussinov  \\
Raymond and Beverly Sackler Faculty of Exact Sciences \\
School of Physics and Astronomy \\
Tel-Aviv University, Ramat-Aviv 69978, Israel \\
E-mail: \email{chj3@post.tau.ac.il},
\email{nussinov@post.tau.ac.il}}

\abstract {We discuss several aspects of astroparticle physics
pertaining to a new model with MeV cold dark matter particles,
which annihilate to electron-positron pairs in a manner yielding
the correct CDM density required today, and explaining the
enhanced electron-positron annihilation line from the center of
the Galaxy. We note that the mass of the vector meson mediating
the annihilations, should exceed the mass of CDM particle, and
comment on possible enhancement due to CDM clustering, on the
detectability of the new CDM, and on particle physics models
incorporating this scenario.}


\preprint{TAUP-2850/07}

\begin{document}
\section{Introduction}
  The possibility that cold dark matter (CDM) consists
of $\sim$MeV scalar particles
 $\chi$'s with cross-sections of $\sim 8-10$\,pb at the freeze-out epoch for annihilating
into electron positron pairs has attracted much attention lately
(see, e.g.,
\cite{Boehm:2003bt,Boehm:2004gt,Borodatchenkova:2005ct,Bouchiat:2004sp,
Fayet:2004bw,Boehm:2003hm,Boehm:2003ha,Beacom:2004pe,Fayet:2006sa,Fayet:2006xd,Fayet:2006sp,
Ahn:2005ck, Hooper:2007jr, Fayet:2007ua} and references therein).
If the annihilation proceeds via an intermediate vector state,
then its rate is proportional to the velocity squared. The reduced
rate today can then  reconcile
 the correct freeze-out  density
 and the excess of the 511 keV annihilation line from
the galactic center \cite{Jean:2003ci},\cite{Knodlseder:2003sv}.

 The new $U(1)'$ abelian gauge interaction whose vector boson
U is responsible for
 the annihilation in the above scenario ties in with
earlier considerations of a light, weakly interacting new vectors
\cite{Fayet:1977vd,Fayet:1977yc,Fayet:1980ad,Fayet:1980rr}.
 While there have been extensive studies of the
prospects and
limitation of the new scenario from both astrophysics
and particle
physics points of view, we found a few points that
to our knowledge
have not been discussed before, and which we present
in the following.

\section{Lower bound on the U mass}

 The lower limit on the prospective CDM particle
mass:
\begin{equation}
   m_\chi  > m_e           
\end{equation}
is  required to kinematically allow the annihilation $\chi \chi
\rightarrow e^+ e^-$. This process, however, is accompanied by the
process $\chi \chi \rightarrow e^+ e^-\gamma$, which is due to
electromagnetic radiative corrections. Thus, a more subtle upper
bound \cite{Beacom:2004pe}:
\begin{equation} \label{eq:m_chi}
   m_\chi < 20 \; {\rm MeV}    
\end{equation}
ensues by requiring that the radiative emission of an extra photon
from the final electron in CDM annihilation will not yield too
many energetic photons violating bounds from EGRET and COMPTEL.

An even more stringent upper bound \cite{Beacom:2005qv}
\begin{eqnarray} \label{eq:bound}
m_\chi<3\; {\rm MeV}
\end{eqnarray}
results from the fact that a small fraction of the energetic
positrons will annihilate to produce energetic gamma rays. Similar
considerations with different assumptions about the ionization
state of the interstellar medium in \cite{Sizun:2006uh} lead to a
more relaxed bound of $m_\chi<7.5\; {\rm MeV}$.\footnote{We would
like to thank John Beacom for bringing this to our attention.}

For the rest of this paper, unless stated otherwise, we will use
the more relaxed upper bound from Eq.(\ref{eq:m_chi}) and set
$m_\chi=10\; {\rm MeV}$.

The required $\chi\chi \rightarrow e^+ e^-$ annihilation cross
section at freeze-out is given by :
\begin{equation}
 v_\chi \sigma_{ann} \simeq \frac{2}{3\pi} v_\chi^2 {G'}^2 m_\chi^2 \sim
 10 \,pb,
\label{freezeout} 
\end{equation}
with $v_\chi^2 \sim 0.05$ at freeze-out (see Eq. (\ref{eq:v_sqr}))
, and fixes
\begin{equation} \label{eq:g_e}
  \frac{g'_e \; g'_\chi}{(m_U^2)} = G'> 10^3\;
  G_F \sim 10^{-2}GeV^{-2},
\label{EqG'GF}
\end{equation}
where $g'_e$ and $g'_\chi$ are the coupling constants of the new
$U$ boson to the electron and the CDM particle. The definition of
$G'$ holds for situations when the total center of mass energy
squared, $s$,
  is lower than $m_U^2$ which is exactly what our new
inequality Eq. (\ref{eq:ineq}) implies in the case of the $\chi
\chi$ annihilations.

 To minimize the effects of $U$ exchange in the well-studied
electron sector one assumes that:
\begin{equation}
 g'_\chi \gg g'_e
\end{equation}
and also that the new $\chi$ particle has no coupling to the $Z$
boson to agree with the $Z$ decay data from LEP. To maintain a
small $g'_e$ and a perturbative picture with $g'_\chi < 1$ one
usually further limits:
\begin{equation}
   m_U < \mathcal{O}(1\, {\rm GeV}).
\end{equation}

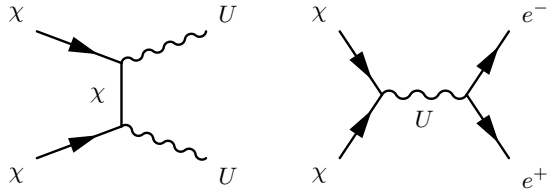
\begin{figure}[!htb]
\center{
\begin{fmffile}{schan}
\fmfframe(10,20)(1,7){
\begin{fmfgraph*}(80,48)
    \fmfleft{i3,i4}
    \fmfright{o3,o4}
    \fmflabel{\footnotesize{$\chi$}}{i3}
    \fmflabel{\footnotesize{$\chi$}}{i4}
    \fmflabel{\footnotesize{$U$}}{o3}\fmflabel{\footnotesize{$U$}}{o4}
    \fmf{fermion}{i3,v3}
    \fmf{fermion}{i4,v4}
    \fmf{plain,label=\footnotesize{$\chi$}}{v3,v4}
    \fmf{boson}{v3,o3}
    \fmf{boson}{v4,o4}
\end{fmfgraph*}}
\fmfframe(30,20)(1,7){
\begin{fmfgraph*}(80,48)
    \fmfleft{i1,i2}
    \fmfright{o1,o2}
    \fmflabel{\footnotesize{$\chi$}}{i1}
    \fmflabel{\footnotesize{$\chi$}}{i2}
    \fmflabel{\footnotesize{$e^+$}}{o1}\fmflabel{\footnotesize{$e^-$}}{o2}
    \fmf{fermion}{i1,v1}
    \fmf{fermion}{i2,v1}
    \fmf{boson,label=\footnotesize{$U$}}{v1,v2}
    \fmf{fermion}{v2,o1}
    \fmf{fermion}{v2,o2}
\end{fmfgraph*}}
\end{fmffile}
\caption{$t$-channel decay into $U$ bosons and $s$-channel decay
into electrons.}}\label{s_chanel}
\end{figure}

We would like to point out that the  CDM scenario of the form
 discussed so far also requires an additional mass bound:
\begin{equation} \label{eq:ineq}
  m_U > m_\chi
\label{mumxi}
\end{equation}
 which becomes stronger when $m_\chi$ approaches the previous,
 upper bound on $m_\chi$ (Eq. (\ref{eq:m_chi})).
 The argument for the new bound is straightforward:
 If  $m_U < m_\chi$, the CDM annihilation into two U bosons
via $\chi$ exchange in the t channel is much faster than
annihilation into $e^+e^-$ (fig. \ref{s_chanel}). The first cross
section is proportional to ${g'_\chi}^4$, and the second to the
much smaller ${g'_\chi}^2 \; {g'_e}^2$.  Also the two U's
annihilation is {\it not} suppressed by the p-wave $v_\chi^2$
factor and completely dominates the annihilations in the halo at
the present time.

\begin{figure}[!htb]
\center{
\begin{fmffile}{box}
 \fmfframe(2,3)(3,5){
\begin{fmfgraph*}(80,48)
    \fmfleft{i1}
    \fmfright{o1,o2,o3}
    \fmflabel{\footnotesize{$U$}}{i1}
    \fmflabel{\footnotesize{$\gamma$}}{o1}\fmflabel{\footnotesize{$\gamma$}}{o2}\fmflabel{\footnotesize{$\gamma$}}{o3}
    \fmf{boson}{i1,v1}
    \fmf{plain,tension=0.5}{v1,v2}
    \fmf{plain,tension=0.1}{v2,v3}
    \fmf{plain,tension=0.1}{v3,v4}
    \fmf{plain,tension=0.5}{v1,v4}
    \fmf{boson}{v2,o1}
    \fmf{boson,tension=0.1}{v3,o2}
    \fmf{boson}{v4,o3}
\end{fmfgraph*}}
\end{fmffile}}
\caption{$U$ decay into three photons} \label{box}
\end{figure}
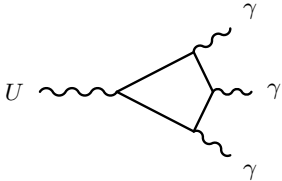

Since the $\chi$'s and the U's stay in thermal equilibrium in the
  early universe until the freeze-out temperature
\begin{eqnarray}
T_F = \frac{m_\chi}{x_F},
\end{eqnarray}
where $x_F \sim \,$18 for a 10\,MeV particle \cite{Fayet:2004bw},
the $\chi$ abundance  will be drastically suppressed relative to
that of the U's:
\begin{equation}
  \frac{n_\chi}{n_U} \sim e^{-(m_\chi - m_U)/T_F} \ll 1.
\end{equation}
The U particles in turn quickly decay and cannot serve as CDM. If
$m_U > 2 m_e$, then the decay $U \rightarrow e^+ e^-$ occurs at
times shorter than a millisecond (even for ${g'_e}^2 < 10^{-16}$).
If  $m_U < 2 m_e$,  U still decays into three photons via an
electron box diagram (fig. \ref{box}). Thus, if the new bound we
suggest is violated we lose the CDM altogether.

  Further difficulties arise from considering the
present day 511 keV positron annihilation line.  For $m_U > 2m_e$
with practically instantaneous $U \rightarrow e^+ e^-$, the rate
of positron production is controlled by $\chi \chi$
 annihilation into $UU$. A substantial cross section for
this annihilation
 is fixed by demanding a correct relic $\chi$ density.

 Since this cross section is {\it not} suppressed by the
$v^2$ p-wave factor,
 the resulting  511 keV  line from the galactic center
will be too strong.

For a p-wave annihilation at the freeze-out epoch
\begin{eqnarray}
    \sigma v = a+bv^2=bv^2 \sim \mathcal{O}(10\,pb),
\end{eqnarray}
which implies $b\sim 200$\,pb for
\begin{eqnarray}\label{eq:v_sqr}
v^2\sim \frac{T_F}{m_\chi} \sim \frac{1}{20}.
\end{eqnarray} For velocities typical to
the center of the galaxy ($v_\chi^2 \sim 10^{-6}$) the
cross-section becomes
\begin{eqnarray}
    \sigma v\sim \mathcal{O}(10^{-4} pb).
\end{eqnarray}

When there is no p-wave suppression the cross-section is
approximately velocity independent, thus dark matte particles
annihilating to two $U'$ bosons will yield a $511$\,keV flux which
is $\sim 10^5$ larger than observed.

  In the case of $m_U < 2 m_e$
 we have no $511$ keV line but rather a continuum from the
$U \rightarrow 3\gamma$ process.

 We may have a consistent cosmology {\it and} present day
annihilation line with
\begin{eqnarray}
    2 m_e < m_U < m_\chi
\end{eqnarray}
if we finely tune $g'_e$ to extremely tiny values ($10^{-28}$)
thereby explaining the annihilation line excess via the ultra-slow
decay of the U particles. This, however results in very large
values of $g'_\chi$ (much larger than one), according to eq.
(\ref{eq:g_e}), rendering perturbative calculations invalid.

\section{CDM clumping and enhanced annihilation rates}

  The rate of annihilation of dark matter is proportional to
   the square of
its local number density.  It scales as $n_\chi(r)^2$
 and generally is enhanced by  DM clustering. Indeed
  the enhancement of the 511 keV  annihilation line from
the galactic center is a basic feature of the present MeV CDM
scenario. The DM density $n_\chi$ is maximal and the resulting
positrons can be brought to rest there.

  However, CDM is clumpy, forming structures on many
smaller scales. Will the rate of mutual annihilation
of CDM particles within any such smaller structure
be enhanced as well?

 We note here that when the annihilation rate is
suppressed by the $v_\chi^2$, p-wave factor, this may {\it not} be
the case.
 This is relevant for the  MeV CDM scenario
discussed here and even more so to certain variants of
LSP (lightest SUSY partner) CDM.
 The smaller mass, yet denser, CDM ``mini-haloes" form
first and larger structures incorporating the previous clusters as
well as some unclustered CDM particles form next. While the
latter, bigger structures have lower densities, they have larger
escape velocities, as required in order to ``bind" unclustered
particles which would escape from the smaller, earlier,
structures.

 The relative velocity of any pair of CDM particles
within the {\it same} mini-halo is of the order of the (small)
escape velocity from this mini-halo. The resulting $v_{rel}^2 \sim
v_{escape}^2$ suppression factor tends to overcome the $n_\chi^2$
enhancement of the annihilation rate of pairs of CDM particles
from the same mini-halo.

  An extreme example illustrating our point is
provided by the $10^{-6}$ solar mass mini-haloes of size $\sim
0.01$ parsec numerically computed \cite{Diemand:2005vz} and
theoretically estimated for the case of heavy LSP CDM particles.

The average density in these mini-halos, $\sim 10 GeV/(cm)^3$, is
$\sim 100$ times larger than the halo density near the center of
the galaxy, so that in the absence of the p-wave suppression
factor the annihilation rate there would be $10^4$ times larger.
However, the escape velocity from the mini-halo
\begin{equation}
    v_{\rm M.H.}=
    \sqrt{G_N \frac {M_{\rm M.H.}}{R_{\rm M.H.}}} \sim
    10^2 \, cm/sec
\end{equation}
is 1/300,000 times smaller than the virial velocity in the
galactic halo
\begin{equation}
    v_{G.H.} \sim 3 \cdot 10^7 \, cm/sec.
\end{equation}

 In the presence of a $v^2$ factor the
rate of annihilation
 of CDM particles from the same mini-halo is smaller
than the
 average annihilation rate in the galactic halo by a
factor:
\begin{equation}
    \left(\frac{v_{M.H.}}{v_{G.H.}}\right)^2
    \left(\frac{\rho_{\chi(M.H.)}}{\rho_{\chi(G.H.)}}\right)^2 = 10^{-7}
\end{equation}
where we used the fact that the CDM mass fixes the
ratio particle number
 and mass densities.
 This feature persists in higher structures though to
lesser and lesser degree.

 Thus, envision a fractal geometric pattern for
successively larger structures.
 The second generation objects just next to
smallest mini-haloes, have masses:
 $M_2 = a \cdot M_1=a \cdot 10^{-6} M_{\odot}$ and radii
$R_2=b \cdot R_1=b \cdot 10^{-2}$ pc; likewise,
 $M_3=a \cdot M_2$; $R_3=b \cdot R_2$, etc., all the way to
$M_N=M_{G.H.} \sim 10^{12}M_{\odot}$ and
 $R_N=R_{G.H.} \sim 50$ kpc. The rates of
 annihilations in successive
 structures also form then a geometric progression
with the ratio $a^3/b^7$ which should be larger than unity,
leading to a gradual increase of same cluster annihilation rate
all the way from the smallest mini-haloes to the galactic halo.
(To account for the fact that the density of the clusters
decreases in each generation, $a/b^3<1$, we need the relation $b^2
< a < b^3$, which should hold for each generation separately).

  It should be emphasized that the annihilation of CDM
particles within a particular cluster with CDM particles coming
from the galactic halo at large are not affected by the
clustering. The relative velocities are then typical galactic halo
velocities and the long time average density seen by the external
particle is that of the galactic halo as a whole as well.

 Thus, in the presence of the p-wave suppression $v_\chi^2$ factor the
only effect of clustering is heterogeneous annihilations occurring
more frequently in the denser regions. If the products are
directly observable energetic gamma rays this can still help in
identifying the CDM.

\section{Can MeV CDM particles manifest in underground
detectors?}

 The underground cryogenic detectors, optimized for
detection of WIMPS of masses
 larger than $\sim$ 100 GeV do not constrain the
proposed MeV CDM scenario:
 The $\chi$ particle can only deposit energies which are
much below the $\mathcal{O}$(keV) detection threshold of the large
bolometric devices.

 Thus, in a collision with an electron the recoil
energy (integrating over all possible incident angels) is
\begin{eqnarray}
    E_{recoil}\sim 2 \cdot \frac{1}{2} m_e \cdot v_\chi^2 \sim \frac{1}{2} eV
\end{eqnarray}
for the standard $3 \cdot 10^7$ cm/(sec) galactic halo virial
velocity.  (If the electron is bound by more than 1/2 eV then the
whole atom recoils with much smaller energies).
 The rate of $\chi-e$ collisions is far
larger than that of the usual
 $M_\chi\sim$ 100 GeV  WIMPs. This stems in part from the
$M_\chi/(m_\chi) \sim  10^4$ larger number
 density and flux of the MeV CDM. Also the  $\chi-e$
 collisions generated by the ``crossed" Feynman diagram
with the U exchanged in the t channel have larger cross sections.
Thus using  Eq. (\ref{freezeout}),
removing the $(v_\chi)^2 \sim 1/20$, p-wave suppression factor and
introducing the kinematical factor $ [m_e/m_\chi]^2 \sim 2.5\cdot
10^{-3}$ yields:
\begin{equation}
  \sigma_{\chi e \rightarrow \chi e}v_\chi \sim \mathcal{O} (0.5\,pb).
\end{equation}
The resulting elastic collision rate is then:
\begin{equation}
n_\chi v_\chi \sigma_{\chi e \rightarrow \chi e} \cdot  n_{\rm
free\; electrons} \sim 100 \;\frac{\rm events}{\rm gr\cdot day}
\end{equation}
where $m_\chi$ = 10 MeV leads to a local CDM density $n_\chi$ of
10 particles/(cm)$^3$ and we estimated  $10^{22} \sim 10^{-2} \;
N_{Avogadro}$ ``free" electrons per gram since only valence
electrons within $1/2$ eV from the Fermi energy surface can
scatter. The relatively large number of expected collisions
suggests using much smaller bolometric devices with
correspondingly lower thresholds. The detection in this case is
not based on nucleon recoil as in the case of heavy WIMPs, but
rather on ionization, which is easier to detect.

Using the more stringent upper bound for $m_\chi$ from Eq.
(\ref{eq:bound}) and setting $m_\chi=3\; {\rm MeV}$ leads to $\sim
4000$ event per gram per day.

Since we rely directly on  $\chi \chi \rightarrow  e^+ e^-$
annihilation rate which is fixed by the present CDM scenario and
the only further theoretical input is crossing, the estimates of
the number of $\chi$ electron collisions are robust, making
efforts to look directly for such CDM most worthwhile.

\section{$U'(1)$ charge conservation, the stability of $\chi$ and
$U-Z$ mixing}

  The most important requirement of any DM candidate
is that it will be stable, or at least have a lifetime exceeding
$t_{Hubble} \sim 5 \cdot 10^{17}$ sec.
 In the present case we have even a stronger bound on
the partial width
 of the putative $\chi \rightarrow e^+ e^-$ decay. Unless
\begin{equation}
 {\Gamma}(\chi \rightarrow e^+ e^-) < 10^{-28} \; sec^{-1}
\end{equation}
the presently decaying $\chi$'s would generate a 511 keV
 annihilation line which exceeds the one observed.

 The dimensionless effective
 $\chi e_L \overline{e_R}$
 coupling $f $ needs then
 to be smaller than $\sim 10^{-24}$ to ensure this bound.
 A nice feature of the present CDM model is that the
conservation of the
 new $U'(1)$ charge could ensure the stability of
$\chi$ if $\chi$ is the lightest particle carrying the specific
$U'(1)$ charge $g'_\chi$.

 The electron also carries a charge $g'_e$ . However,
$U'(1)$ charges need not be quantized and, indeed, phenomenology
demands that $g'_\chi \gg g'_e$
 which automatically guarantees the stability of
CDM.
 While there must be a spontaneous breaking of the
local $U'(1)$ to provide mass---via a Higgs mechanism---to the
associated gauge boson U, no renormalizable trilinear vertex can
violate $U'(1)$.

 This last feature is a source of considerable
difficulty with models in which the new $U'(1)$ is not vectorial.
An example of the latter is purely $U'(1)_R$  ``minimalistic"
model \cite{Borodatchenkova:2005ct} where the doublet $(e_L,
\nu_L)$ and all quarks and leptons (other than $e_R$) are assumed
to be $U'(1)$ neutral. We recall the following mixing argument
\cite{Fayet:1986rh,Fayet:1989mq,Fayet:1990wx,Fayet:2007ua}:

 Any Higgs couplings $g_e H e_L \overline{e_R}$ implies that
the Higgs in question -- the standard model Higgs or another Higgs
doublet $\tilde{H}$ must also carry a $U'(1)$ charge $g' =
g'_{e_R}$ to match that of $e_R$.  An additional Higgs is required
in supersymmetric theories and could, in principle, also provide
the U mass (the single Higgs $H^0$ of the standard model accounts
for only one longitudinal degree of freedom and
 makes only one gauge boson, namely, the $Z^0$, massive).

 Since the weak doublet $\tilde{H}$ couples to both Z and U
bosons we find a $2 \times 2$ (Z,U) mass$^2$ matrix:
\begin{eqnarray}
M^2=\left(\begin{array}{cc} g^2 v^2 + {g'}^2 \tilde{v}^2\ \  & gg'
v\tilde{v} \\
gg'v\tilde{v}\ \   & {g'}^2 \tilde{v}^2
\end{array} \right)
\end{eqnarray}

\noindent where $g$ and $v = 246$\,GeV are the ordinary weak
coupling and Higgs vacuum expectation value. Diagonalizing the
mass matrix yields a physical Z with essentially the same mass of
$gv$ and a tiny irrelevant admixture of U.

 The important point is that the physical U is now:
\begin{equation}
 U_{Phys} = U + [\tilde{v}g'/vg] \cdot Z \equiv U+{\epsilon}Z.
\end{equation}
and has a ``see-saw" like mass:
\begin{equation}
 m_{Phys}(U) = \frac{{g'}^2 \tilde{v}^2}{m_Z}.
\end{equation}

The Z ``masquerading" as the light U contributes too much to
processes such as $\nu -e$ scattering unless:
 \begin{equation}
\epsilon  = \frac{\tilde{v}g'}{vg} < \frac{m_U^2}{m_Z^2} \cdot
10^{-2},
\end{equation}
where the last factor is roughly the precision to which
the standard
model Z exchange correctly describes low energy
neutrino electron scattering.

 If we have only one Higgs (in which case, all
right-handed fermions have $ U'(1)$ charges equal to $g'$) the
last upper bound (and others coming from considering atomic parity
violating experiments \cite{Bouchiat:2004sp}) would too strongly
limit $g'$ making the  CDM model untenable. Thus we need to take
$\tilde{H} \neq H$ and
\begin{eqnarray}
\tilde{v} \sim GeV \ll \left< H \right> =v = 246 GeV.
\end{eqnarray}
Also, since $g'< 10^{-2}$, the above U mass, is too small
violating our
bound of Eq. (\ref{mumxi}). 

 To generate the required U mass we need then an
additional new $U'(1)$ carrying boson taken for simplicity as an
$SU(2)_L$ singlet, $\phi$, with a vev $\left< \phi \right>$ such
that $m(U) = g'_\phi \left< \phi \right>$. We cannot use the CDM
particle $\chi$ itself for this purpose since the Higgs mechanism
leaves us with just one real, self charge conjugate, scalar which
will not couple to the vector U.

 The need to have $U'(1)$ charged Higgs doublets and
ensuing $Z-U$ mixing
 is avoided if $U'(1)$ is vectorial. The right- and
left-handed fermions carry then the
 same $U'(1)$ charges and the Higgs stays U'(1) neutral.

\section{Anomalies and ``Hard" $F_{(\mu \nu)}^{em} F'_{(\mu \nu)}$
mixing}

 The cancelation of all triangular
$V_\alpha V_\beta A_\gamma$ axial anomalies is required in any
gauge theory.
  For $ SU(3) \times SU(2) \times U(1)$  with the standard assignment
of hypercharge
and weak isospin, all anomalies cancel in each
generation
separately. (Indeed, with one extra right-handed
neutrino the 16 fermions in each
 generation form a representation of the anomaly free
$SO(10) $ group.)

 If the new $U'(1)$ is not purely vectorial, many
additional $U'(1)^3, \; U'(1)U_Y^2, \; U'(1)U_{em}^2$, etc.,
anomalies arise; the cancelation of which requiring  extra,
electrically charged fermions. For vectorial $U'(1)$ we have only
$U'(1)^2 \; U_{SM}$ type anomalies which
 cancel if $U'(1)$ couples equally to all fermions or
like $B-L$.

 In principle we need also to worry about ``hard"
$U-\gamma$ mixing generated via the electron loop. Unless the
logarithmically divergent part is canceled, an $F \cdot F'$ mixing
would be generated. This $U-\gamma$ mixing makes mini-charged
$\chi$'s and CDM, with many additional problems. A mixing of two
U(1) gauge groups is avoided if the latter are embeddable in a
large Lie group with the quarks and leptons forming an irreducible
representation of
 the latter. The cancelation of the divergent parts
of all mixings, namely,
 $Tr(Q_\alpha  Q_\beta) = \delta_{\alpha \beta}$
is then automatic.

 It will also happen here if, as suggested in \cite{Bouchiat:2004sp}, the
$U'(1)$ couples universally to all fermions as then the
cancelation conditions   $Tr(Q_{em})=0$, etc., are all satisfied.

\section{Can the new U mesons be searched for in hadronic
machines?}

 If the new U vector meson couples equally to quarks
and leptons (in a given generation) as the above arguments may
suggest, then low energy high current proton beams on a fixed
target set-up may provide ideal hunting grounds for the new U
vector. If $m_U < m_\pi$ we can
 work below pion threshold and look for, say,
$pp \rightarrow pp \; U$ with
 the U decaying into $e^+ e^-$.

Further, if $m_U < 2 m_\chi$, the invisible decay mode,
$U\rightarrow \chi\chi$, which would otherwise dominate by $\sim
10^6$ factor or more is absent. Altogether this set-up would then
have much higher statistics and far smaller radiative QED
``trident" diagram contributions than in $e^+ e^-$ colliders or
$e^- p$ experiments.

\section{Is there a consistent SM $\times$ U$'$(1) model for the
MeV CDM scenario?}

 The absence of dangerous $U-Z$ mixing and anomalies
strongly suggest  $U'(1)_V$.
 However, in that case  an $SU(3) \times SU(2)_L \times U(1) \times U'(1)_V$
implies that the U'(1) vector boson couples to $e_R, e_L$ and the
other member of the $SU(2)_L$ doublet $\nu(e)_L$ with the same
coupling
\begin{eqnarray}
g'_e=g'_\nu \equiv g',
\end{eqnarray}
which, as pointed out by several authors, entails various
difficulties.

At energies
\begin{eqnarray}
s=2m_e \cdot E_\nu \sim (10 MeV)^2
\end{eqnarray}
neutrino-electron scattering has been studied with precision in
accelerators. The U exchange amplitude  $\sim g'^2/(m_U)^2$ (for
an assumed $s < m_U^2$) should be smaller than $\sim $ 1\% of the
weak amplitude $G_F$ (the precision level). Comparing this with
our required $g'_e g'_\chi/(m_U)^2 > 10^3 G_F$ (Eq. (\ref{EqG'GF})
above), this implies that $g'_e/g'_\chi < 10^{-5}$.

The authors of ref. \cite{Fayet:2006sa} pointed out also a
qualitative astrophysical argument which tends to strongly exclude
a vectorial $U'(1)_V$.
 The argument runs as follows: The  bound
\begin{eqnarray}
m_\chi < 20 MeV
\end{eqnarray}
 implies that these CDM particles  will be
abundantly produced in
 a supernova collapse. The scattering cross sections of
$\chi$ off both electrons and neutrinos are (at least) $\sim 10^6$
times larger than the typical weak $(G_F^2 E^2)$ cross sections at
these energies.  Hence, the $\chi$'s will be trapped by the
electrons and, in turn, trap the neutrinos. This yields longer
cooling times and lower energies of the emitted neutrinos than
what was observed for supernova SN 1987(a).

 Our bound $m_U > m_\chi$ only strengthens this
argument: as we push to higher $m_\chi$ to suppress the number of
these problematic particles
 in the core of the supernova, $m_U$ becomes larger
and possible
reductions
 of the $e-\chi$ and $\nu-\chi$ cross sections by
the U propagator for
 energies $E > m_U$ do not occur in the supernova
core.

\section{Could the new U boson be associated with  a
``Horizontal" gauge group?}

 Optimally the new U boson and new $U'(1)$ would match
some expectations of specific new physics beyond the
standard model.
 The fact that Fayet has anticipated a new light U and
associated
$U'(1)_V$ some time ago in a particular SUSY setting is
very intriguing.

 Here we would like to comment on the possibility that
the new U boson  is associated with a gauged horizontal symmetry
group. Indeed, the dynamics of the three families and their mass
pattern is an outstanding puzzle which has prompted considering
Horizontal groups (H.G.'s - see for instance
\cite{Davidson:1979nt}). These groups must be spontaneously
broken. For global H.G. the resulting massless Goldston
``Familons" could manifest in $K \rightarrow \pi$ + ``missing zero
mass particles" or similar muon to electron decays. This then
implies very severe bounds $(F_H > 10^{10}$ GeV) on the symmetry
breaking scale.

The situation is better when we have gauged H.G. with a Higgs
mechanism providing masses to the horizontal gauge bosons.
Horizontal gauge couplings similar to the above $g'_e \sim
10^{-4}$ and masses $\sim m_{Z_H} \sim 100$ GeV for the
intergenerational horizontal vector bosons are consistent with the
existing limits on $\mu \rightarrow 3 e$ and $K \rightarrow \pi +
2\nu$.
 We assume that the light U in the present CDM
scenario corresponds to a more weakly broken $U'(1)$ which couples
only to the first generation quarks and leptons with MeV - $10$
MeV masses of order of the U mass itself.

 Note that all the anomaly and other cancelation
discussed above occur for each generation separately
and are consistent with U coupling to the first
generation fermions only.

 Amusingly  also the difficulty with super-nova data is partially
alleviated if only the electron and its neutrino couple to
$U'(1)$. The $\mu$ and $\tau$ type neutrino would then normally
escape had it not been for the, say, annihilation process $\nu_\mu
+ \overline{\nu_\mu} \rightarrow Z \rightarrow \nu_e + \overline
{\nu_e}$. Since the $\mu,\tau$ neutrinos keep escaping this
reaction will proceed mainly in the reverse direction and the
extended delay largely avoided.
 Also the relative precision of measurement of
electron neutrino scattering is lower than that of the $\mu$
neutrino, and ensuing limits are less stringent.

\section{Summary and Conclusions}

 In the above we have made various comments pertaining
to the new MeV CDM
 scenario and its possible embedding in a particle
physics model.

 Our first result, $m_U > m_\chi$ becomes
particularly useful in
 conjunction with other restrictions on the model.
 We next pointed out the general amusing results that
when the p-wave $v^2$ suppression factor is present, the overall
rate of annihilations in small CDM mini-halos  need not be
enhanced.

 We noted the likely direct detection of the MeV CDM
by small bolometric underground devices, using CDM---electron
scattering---the cross section of which follows in an almost
model-independent way from the crossed annihilation process which
is the basic ingredient of the MeV CDM scenario. Also the easier
detection of light U boson in proton fixed target experiments was
noted.

 Finally we commented on some more theoretical aspects
of particle physics models for the scenario, and, in
particular, on  possible connections with horizontal
gauged symmetries.

 The new MeV CDM model puts strong demands on any
underlying particle
 physics model. We need a new light  scalar $\chi$
when to date no
 elementary scalars have been observed at any mass.
 We also need a new $U'(1)$ with very large  ratios  of
the $U'(1)$ charges   of the $\chi$ and electron when all known
electric charges are quantized to a $10^{-21}$ accuracy.

 All that notwithstanding, we believe that one should
try as hard as
 possible to experimentally refute this scenario, or
verify it and
 the abundant new physics which necessarily attends
it.

\bibliographystyle{JHEP}

\bibliography{C:/temp/MeV}

\end{document}